\documentclass[11pt, copyright, logo, secondlogo]{tex/googledeepmind}

\makeatletter
\RequirePackage{authblk} 
\renewcommand\AB@affilnote[1]{}%
\renewcommand\AB@authnote[1]{}%
\makeatother

\usepackage{bbm}

\usepackage{amsmath,amsfonts,bm}

\def\eqref#1{equation~\ref{#1}}

\def\1{\bm{1}}

\DeclareMathAlphabet{\mathsfit}{\encodingdefault}{\sfdefault}{m}{sl}
\SetMathAlphabet{\mathsfit}{bold}{\encodingdefault}{\sfdefault}{bx}{n}

\usepackage[authoryear, sort&compress, round]{natbib}
\usepackage{hyperref}
\usepackage{url}
\usepackage{algorithm}
\usepackage{booktabs}
\usepackage{pifont}
\usepackage{graphicx} 
\usepackage{caption}
\usepackage{amssymb}
\usepackage{tcolorbox}
\tcbuselibrary{listings,skins,breakable}
\usepackage{wrapfig}
\usepackage[dvipsnames]{xcolor}
\usepackage{comment}
\usepackage{float}
\usepackage{subcaption}
\usepackage{enumitem}
\usepackage{soul}
\usepackage[frozencache]{minted}
\usepackage{algorithm}
\usepackage{algorithmic}
\usepackage{booktabs}
\usepackage{multirow}
\usepackage{pifont}
\newcommand{\cmark}{\ding{51}}
\usepackage{enumitem}
\setlist[itemize]{topsep=0pt, itemsep=-3pt}
\definecolor{codebg}{rgb}{0.95,0.97,1.0}
\setminted{bgcolor=codebg}

\definecolor{basegptfive}{RGB}{153, 204, 255}     
\definecolor{claudesonnet}{RGB}{255, 99, 71}      
\definecolor{claudeopus}{RGB}{255, 193, 7}        
\definecolor{gemini}{RGB}{144, 238, 144}          
\definecolor{gptfiveRL}{RGB}{34, 139, 34}         
\definecolor{grokfour}{RGB}{64, 224, 208}         
\definecolor{grokcode}{RGB}{147, 112, 219}        
\definecolor{makora}{RGB}{25, 25, 112}            

\title{Fine-Tuning GPT-5 for GPU Kernel Generation}

\author{Ali Tehrani}
\author{Yahya Emara}
\author{Essam Wissam}
\author{Wojciech Paluch}
\author{Waleed Atallah}
\author{\L ukasz Dudziak}
\author{Mohamed S. Abdelfattah}

\affil{Makora}

\renewcommand{\comment}[1]{}

\begin{abstract}
Developing efficient GPU kernels is essential for scaling modern AI systems, yet it remains a complex task due to intricate hardware architectures and the need for specialized optimization expertise. 
Although Large Language Models (LLMs) demonstrate strong capabilities in general sequential code generation, they face significant challenges in GPU code generation because of the scarcity of high-quality labeled training data, compiler biases when generating synthetic solutions, and limited generalization across hardware generations. 
This precludes supervised fine-tuning (SFT) as a scalable methodology for improving current LLMs.
In contrast, reinforcement learning (RL) offers a data-efficient and adaptive alternative but requires access to relevant tools, careful selection of training problems, and a robust evaluation environment. 
We present Makora’s environment and tools for reinforcement learning fine-tuning of frontier models and report our results from fine tuning GPT-5 for Triton code generation.
In the single-attempt setting, our fine-tuned model improves kernel correctness from 43.7\% to 77.0\% (+33.3 percentage points) and increases the fraction of problems outperforming TorchInductor from 14.8\% to 21.8\% (+7 percentage points) compared to baseline GPT-5, while exceeding prior state-of-the-art models on \texttt{KernelBench}.
When integrated into a full coding agent, it is able to solve up to 97.4\% of problems in an expanded \texttt{KernelBench} suite, outperforming the PyTorch TorchInductor compiler on 72.9\% of problems with a geometric mean speedup of 2.12$\times$.
Our work demonstrates that targeted post-training with reinforcement learning can unlock LLM capabilities in highly specialized technical domains where traditional supervised learning is limited by data availability, opening new pathways for AI-assisted accelerator programming.
\end{abstract}

\begin{document}
\maketitle

\section{Introduction}

Accelerators such as GPUs form the computational backbone of modern AI workloads. Their ability to execute massive numbers of parallel operations has enabled the training and inference of large language models at unprecedented scales. This shift toward accelerator-centric computing is rising rapidly with millions of GPUs being deployed in data centers in 2025 alone~\citep{NVIDIA2025OpenAI, AMD2025OpenAI}. This reflects a broader architectural transition from CPU-centric to accelerator-centric systems, as domain-specific hardware becomes the practical path to continued performance gains. The trend is driven both by increasing AI infrastructure demand and by limits of traditional CPU scaling. With the end of Moore's Law \citep{theis2017end} and the breakdown of Dennard scaling, single-threaded CPU performance has plateaued, making parallel accelerators essential for continued performance growth across the entire computing stack \citep{hennessy2019new}.

However, programming GPUs, commonly referred to as kernel development, remains a challenging and time consuming task that requires significant expertise and careful performance tuning.
Developers must navigate complex memory hierarchies (shared vs. global memory), manage thread synchronization and scheduling, optimize instruction-level parallelism, and handle data movement across heterogeneous compute units. As a result, efficient kernel development remains concentrated among a small group of highly specialized experts and the majority of optimized kernels are proprietary and unavailable to the broader research community. This expertise gap creates a critical bottleneck: as accelerators become widely deployed at scale, the ability to efficiently utilize their memory hierarchies and data movement patterns becomes the limiting factor in realizing their computational potential. Kernel programming expertise remains limited, posing a significant challenge for modern language models. Beyond general-purpose libraries, these models depend on carefully optimized, application-specific kernels that reduce memory traffic and leverage hardware parallelism to achieve competitive performance and efficiency \citep{dao2022flashattentionfastmemoryefficientexact}. 

\subsection{Limitations of Supervised Fine Tuning for GPU Kernel Generation}

Recent advances in large language models have shown strong capabilities in software engineering tasks, including code generation~\citep{chen2021evaluatinglargelanguagemodels, claudecode2024}, program synthesis~\citep{austin2021programsynthesislargelanguage}, and automated bug repair~\citep{Xia_2023, bouzenia2024repairagentautonomousllmbasedagent}. These results also motivate applications of large language models to accelerator programming, with recent work exploring GPU kernel generation through agentic systems~\citep{lange2025robustagenticcudakernel}, reinforcement learning~\citep{baronio2025kevinmultiturnrlgenerating}, and evolutionary optimization~\citep{guo2025evoengineermasteringautomatedcuda}. Benchmarks such as \texttt{KernelBench}~\citep{ouyang2025kernelbenchllmswriteefficient} and datasets such as \texttt{KernelBook}~\citep{kernelbook} have been introduced to support evaluation and training in this domain. 
Although these efforts show promise, substantial progress is still required before large language models can reliably generate production quality high performance GPU kernels. 
Moreover, due to numerous challenges in this domain, supervised training is significantly more difficult than in other programming settings as we describe in more detail below.
Specifically, GPU kernel generation presents a set of challenges that differentiate it from general purpose code generation. These challenges can be broadly grouped into four categories: data scarcity and quality, limitations of compiler generated synthetic data, the insufficiency of functional correctness as an optimization objective, and the exponential growth of the optimization space.

\paragraph{Data scarcity and quality.} High performance GPU kernels require deep hardware specific knowledge and are typically proprietary. Public repositories consist mainly of educational examples or suboptimal implementations, which limits the construction of large scale and high quality training datasets for supervised learning and further restricts the availability of labeled datasets. Unlike general Python or Java code, where millions of examples are available, production grade kernel implementations number in the thousands at best. For example the Python subset of the Stack v2 dataset~\citep{lozhkov2024starcoder} contains 96,448,523 samples, whereas the KernelBook~\citep{kernelbook} dataset which consists of PyTorch and synthetic Triton code only contains 18,162 examples.

\paragraph{Limitations of compiler-generated synthetic data.} An alternative to collecting expert-written kernels is to generate synthetic training data using compilers (e.g., \texttt{TorchInductor}, \texttt{XLA}, \texttt{TVM}). Although the data is abundant, compiler-generated kernels suffer from several critical limitations that constrain the performance of models trained on such data:
\begin{itemize}[leftmargin=*,nosep]
    \item \textit{Performance ceiling}: Compiler-generated kernels reflect the optimization heuristics encoded in the compiler itself. Models trained on this data learn to reproduce compiler strategies rather than discover novel optimizations, creating an inherent upper bound on achievable performance that cannot exceed the compiler's own capabilities.
    \item \textit{Compiler boilerplate and artifacts}: Compilers often generate code with extensive boilerplate, intermediate variables, and code patterns specific to their internal representation. This introduces noise into the training data and teaches models to reproduce compiler-specific idioms rather than clean, readable, and maintainable kernel code.
    \item \textit{Use of internal compiler libraries}: Compiler-generated code frequently relies on internal runtime libraries, intrinsics, or undocumented APIs that are not accessible or portable outside the compiler's execution environment. Models trained on such code may generate kernels that depend on these internal libraries, limiting their practical utility.
    \item \textit{Lack of readability and maintainability}: Code generated by compilers prioritizes correctness and performance over human readability. The resulting kernels often lack clear structure, meaningful variable names, and explanatory comments, making them poor training examples for teaching models to write maintainable production code.
\end{itemize}

\paragraph{Correctness is insufficient.} A functionally correct kernel that produces the right output may still be orders of magnitude slower than an optimized implementation. Even when ground-truth optimized kernels are available, the optimization strategies are hardware-specific and may not generalize across GPU architectures, such as NVIDIA Hopper versus Blackwell. Furthermore, specific devices have different hardware support even within the same hardware architecture generation. For example, within the Blackwell family of devices, the desktop-grade RTX 5090 does not include Tensor Memory Accelerators (TMA), while the server-grade B200 does. These discrepancies, along with differences in memory technology, speed, low-precision support, and distributed computing APIs, severely hamper performance portability across devices.

\paragraph{Exponential optimization space.} Kernel optimization involves discrete choices (tiling strategies, memory layouts, vectorization patterns) that interact in complex, non-linear ways. The space of possible implementations grows exponentially with problem complexity, making it impractical to enumerate all variants for supervised learning.

\subsection{Reinforcement Learning for Kernel Generation}

As large language models (LLMs) approach the data saturation limit for pretraining~\citep{villalobos2024position}, post-training techniques such as reinforcement learning (RL) have gained significant attention. We reformulate the task of kernel generation as a reinforcement learning problem, where the model learns from execution-based feedback. This approach alleviates the dependency on large quantities of high-quality training data while enabling the model to explore the optimization space in a flexible way and discover novel kernel implementations.

\subsubsection{Reinforcement Learning from Verifiable Rewards (RLVR)}
\label{subsubsec:rlvr}

Traditional reinforcement learning from human feedback \citep[RLHF;][]{christiano2023deepreinforcementlearninghuman, ouyang2022traininglanguagemodelsfollow, dai2024safe, kaufmann2025a, lambert2025reinforcementlearninghumanfeedback} relies on learned reward models trained on human preferences, which can be subjective, expensive to collect, and prone to reward hacking~\citep{gao2022scalinglawsrewardmodel}. In contrast, Reinforcement Learning from Verifiable Rewards (RLVR)~\citep{shao2024deepseekmathpushinglimitsmathematical, deepseekai2025deepseekr1incentivizingreasoningcapability} uses deterministic, rule-based reward functions that provide unambiguous binary or continuous feedback based on objectively verifiable criteria~\citep{lambert2025tulu3pushingfrontiers, wen2025reinforcementlearningverifiablerewards}.

For kernel generation, RLVR is particularly well-suited because correctness and performance can be verified automatically through compilation, testing, and benchmarking. Given a kernel generation problem $p$ (consisting of a natural language prompt and a reference PyTorch implementation), the policy model $\pi_\theta$ generates a candidate kernel $k \sim \pi_\theta(\cdot \mid p)$. We define a verifier $\mathcal{V}$ that evaluates the kernel and produces a scalar reward:

\begin{equation}
\mathcal{V}(k,p)=
\begin{cases}
0 & \text{if } k \text{ fails to compile or produces incorrect output} \\
\sigma\!\left(r_{\text{raw}}(k,p)-\delta\right) & \text{if } k \text{ is functionally correct}
\end{cases}
\end{equation}

where $\sigma(x)=\frac{1}{1+e^{-x}}$ is the logistic function, $\delta$ is a shift parameter, and the unnormalized reward $r_{\text{raw}}$ is defined as
\begin{equation}
r_{\text{raw}}(k,p)=\mathbb{1}\!\left[\text{validated}(k,p)\right]+\max\!\left(0,\text{speedup}(k,p)\right).
\end{equation}
Here, $\mathbb{1}[\text{validated}(k,p)]=1$ if the kernel produces correct output and $0$ otherwise. The speedup is measured relative to a baseline implementation, for example \texttt{torch.compile} with the default TorchInductor backend,
\begin{equation}
\text{speedup}(k,p)=\frac{t_{\text{baseline}}(p)}{t_k(p)}.
\end{equation}

The shift parameter $\delta$ controls the relative influence of correctness and performance by determining how much performance improvement is required beyond functional correctness to obtain a high reward. Larger values of $\delta$ require substantially higher speedups for the reward to saturate, thereby leaving more room for performance driven differentiation among correct kernels. Conversely, smaller values of $\delta$ cause the reward to saturate closer to the correctness threshold, reducing the marginal incentive for additional performance improvements and placing greater emphasis on functional validation.
We set the shift to $1.8$ by default, meaning that a correct kernel with performance comparable to \texttt{TorchInductor} receives a reward close to half of the maximum value $0.5$, while kernels with higher performance obtain correspondingly larger rewards approaching $1$.

Recent work demonstrates that RLVR not only improves final outcomes but also implicitly incentivizes correct reasoning processes, even when rewards are based solely on outcome correctness~\citep{wen2025reinforcementlearningverifiablerewards}.
RLVR's deterministic rewards eliminate the ambiguity and bias inherent in human-annotated preferences. However, RLVR is not immune to reward hacking, as models can still exploit loopholes in the verification process to achieve high rewards without genuinely solving the task. We address these challenges through static code analysis, an LLM as a hack-judge, and careful reward design, which we discuss in detail in Section~\ref{sec:reward_hack}.

\paragraph{The Role of Strong Base Models.} A fundamental prerequisite for successful RLVR training is that the base model must already possess basic competency in the target domain. If the model cannot generate syntactically valid code or understand the basic structure of Triton kernels, nearly all generated samples will fail compilation, resulting in uniformly zero rewards. In such cases, the model receives no learning signal and cannot improve through RL. This cold start problem is particularly important in specialized domains like kernel programming, where even syntactically correct code requires deep understanding of hardware primitives, memory hierarchies, and parallelization patterns.

This is why we employ GPT-5 \citep{OpenAI2025GPT5}, a highly capable base model that already demonstrates strong code generation abilities and understanding of GPU programming concepts. GPT-5 has existing knowledge of Triton syntax and parallel programming patterns that ensures that a substantial fraction of initial samples compile successfully and produce non-zero rewards, which provides the gradient signal necessary for RL to drive improvements. Models without this foundation would require extensive supervised pre-training or fine-tuning on kernel code before RL could be effective, which would significantly increase the data requirements that RL was meant to alleviate. The choice of base model is therefore not only a matter of initial quality but also a prerequisite for the feasibility of RL training itself.
We experimented with RL training on smaller open models such as Qwen-4B, Qwen-8B and Qwen-32B and observed that the reward value quickly plateaus, which suggests that the model is able to generate kernels for a limited number of problems and can not obtain further improvements. This behavior has also been observed by other works \citep{baronio2025kevinmultiturnrlgenerating}.

\subsubsection{Scaling RL Training for Kernel Generation}

Scaling Reinforcement Fine Tuning (RFT) to kernel generation is non-trivial and requires attention to several factors. First, a curated training dataset that spans diverse computational patterns. Second, a robust evaluation infrastructure that can efficiently execute generated code and provide meaningful reward signals, mechanisms to prevent reward hacking where the model exploits loopholes in the reward function, and tools that enable the model to iteratively debug and refine kernels.

The experimental results demonstrate fine-tuning GPT-5 for Triton \citep{tillet2019triton} code generation achieves superior performance compared to existing large language models, establishing new state-of-the-art results. Specifically, with three attempts, our fine-tuned model successfully generates 221 functionally correct kernels out of 264 total test cases. Furthermore, beyond functional correctness, the model achieves a new benchmark in geometric mean speedup over \texttt{TorchInductor} on an expanded KernelBench test set~\citep{ouyang2025kernelbenchllmswriteefficient}, highlighting its capability to produce not only correct but also highly optimized kernels for modern accelerator architectures.

When integrated within MakoraGenerate agent, the agent achieves 2.12$\times$ geometric mean speedup and solves almost all \texttt{KernelBench} tasks (up to 97.4\% correctness), outperforming \texttt{TorchInductor} on 72.9\% of problems.

Our main contributions are as follows:
\begin{enumerate}
    \item \textbf{Dataset}: We introduce a dataset curation methodology for reinforcement fine tuning, encompassing a diverse set of kernel generation problems that span computational domains such as matrix multiplication, convolution, reduction, elementwise operations, common fusion patterns, and common AI primitives.
    \item \textbf{Robust Evaluation Infrastructure}: We introduce a scalable evaluation environment that addresses key challenges in RL for code generation:
    \begin{itemize}
        \item Multi-turn refinement through tool-augmented generation, enabling iterative debugging within single trajectories
        \item Comprehensive reward hacking prevention via static reachability analysis tools and LLM based detection of hacks
        \item Efficient evaluation through caching and canonicalization of kernel executions
        \item Distributed evaluation backend supporting parallel kernel benchmarking on H100 GPUs
    \end{itemize}
    \item \textbf{Strong empirical results}: We achieve state-of-the-art performance on Triton GPU kernel generation, surpassing all prior work to our knowledge. In the single-attempt setting, our fine-tuned model attains 77.0\% functional correctness, produces kernels that outperform \texttt{TorchInductor} on 21.8\% of benchmark problems, and achieves a geometric mean speedup of $0.81\times$ relative to \texttt{TorchInductor}, demonstrating both high reliability and competitive performance.
\end{enumerate}

\section{Methodology}

In this section, we describe the fine-tuning process of GPT-5 for Triton kernel generation. We begin by outlining the dataset creation procedure, including the construction of the training and testing sets. Next, we introduce the \texttt{Makora} environment, tools and grader components.

\begin{figure}
    \centering
    \includegraphics[width=1\textwidth]{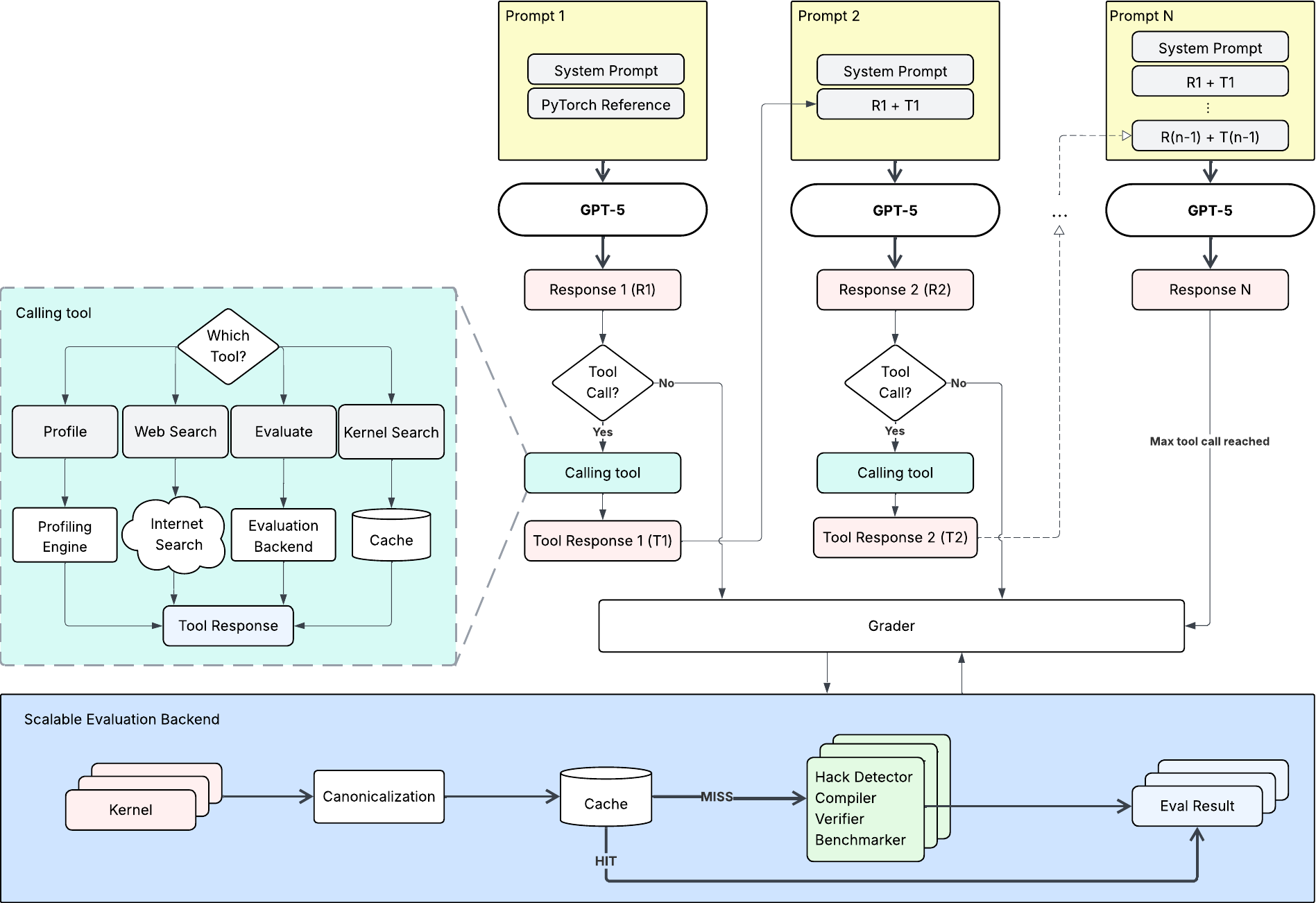}    \caption{Reinforcement Learning fine tuning with Makora's evaluation infrastructure and tools.
    }
    \label{fig:RFT_workflow}
\end{figure}

Figure \ref{fig:RFT_workflow} illustrates the overall reinforcement learning workflow for kernel generation. Each training episode consists of a multi step interaction between the model and external tools. Given a system prompt and a problem specific reference in PyTorch, the model produces an initial kernel generation attempt. At each step, the model may choose to invoke an external tool, including kernel evaluation, database lookup, or web search, to gather additional information relevant to the task. Tool outputs are appended to the model context and used to guide subsequent generations, enabling iterative refinement until a final response is produced or a maximum number of tool calls is reached. The generated kernel is then submitted to a scalable evaluation backend that performs compilation, verification, correctness checking, and benchmarking, with caching to avoid redundant evaluations. The resulting evaluation score is aggregated by a grader and used as the reinforcement learning reward to update the model parameters, directly linking kernel execution outcomes to policy optimization.

\subsection{Dataset}

\begin{figure}
    \centering
    \includegraphics[width=1\linewidth]{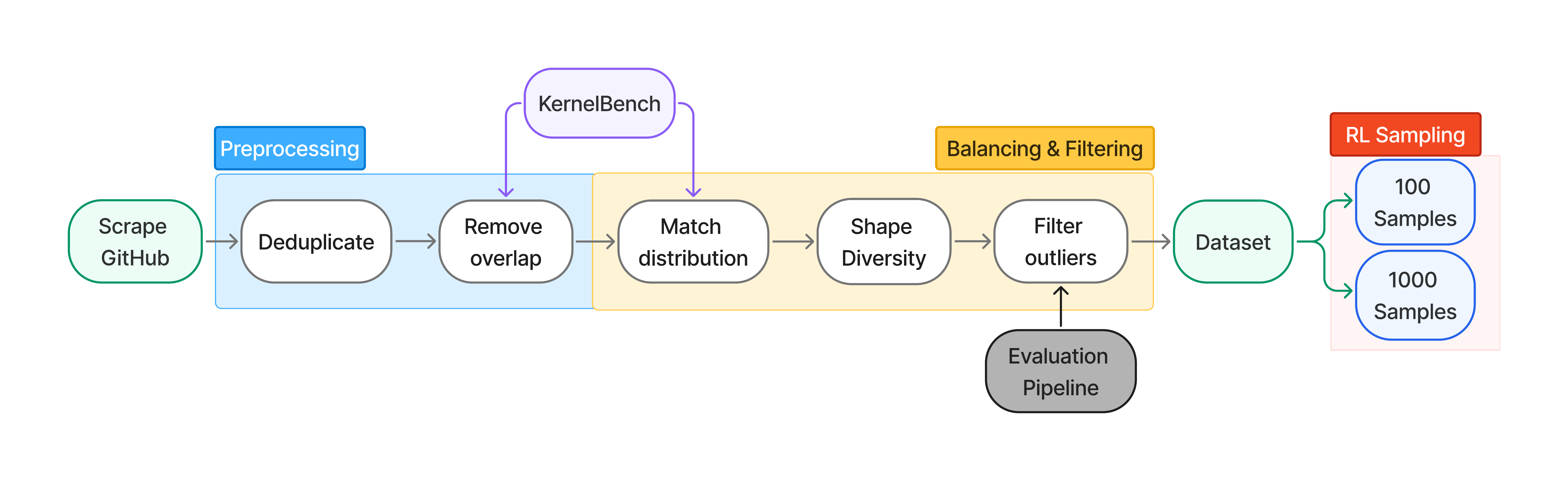}
    \caption{Workflow of dataset processing and cleaning.}
    \label{fig:dataset_cleaning_process}
\end{figure}

Our dataset construction process is designed to produce a diverse, balanced, and high-quality training set for GPU kernel generation. We crawl PyTorch examples from public repositories and apply targeted curation techniques to ensure coverage across problem difficulty and computational patterns while preventing data leakage.
Figure~\ref{fig:dataset_cleaning_process} illustrates our complete data preparation pipeline.

\subsubsection{Base Dataset Construction}

We construct the training and validation sets from two sources.
The validation set consists of examples from KernelBench.
The training set is derived from PyTorch code collected from public repositories and undergoes extensive multi step data quality filtering.
After collection, we apply a sequence of data refinement procedures, including deduplication, dataset decontamination, shape diversification, and runtime based filtering, which we describe below.

\subsubsection{Deduplication Pipeline}

To ensure no overlap between our training set and test set, we implement a two-stage deduplication pipeline:

\paragraph{Embedding-based semantic deduplication.} We embed each PyTorch example in our trainset using \texttt{jina-embeddings-v3}~\citep{sturua2024jinaembeddingsv3multilingualembeddingstask}, a state-of-the-art code embedding model. We apply the same embedding process to all \texttt{KernelBench} examples. For each train sample, we compute the L2 distance to all \texttt{KernelBench} samples in the embedding space. We remove any sample where:

\begin{equation}
\min_{b \in \text{KernelBench}} \|E(\text{code}_k) - E(\text{code}_b)\|_2 < \tau_{\text{embed}}
\end{equation}

where $E(\cdot)$ is the embedding function and $\tau_{\text{embed}} = 0.45$ (corresponding to approximately 90\% cosine similarity). This threshold was chosen through empirical analysis to balance between removing true duplicates and retaining semantically similar but distinct problems.
This step removes primarily problems that are direct copies or minor variations of \texttt{KernelBench} problems.

\paragraph{Syntactic deduplication.} Beyond semantic similarity, we perform exact and near-exact deduplication by computing token-level Jaccard similarity. For each pair of code samples $(c_i, c_j)$, we compute:

\begin{equation}
J(c_i, c_j) = \frac{|T(c_i) \cap T(c_j)|}{|T(c_i) \cup T(c_j)|}
\end{equation}

where $T(\cdot)$ tokenizes code into Python tokens using the \texttt{tokenize} module. We remove pairs where $J(c_i, c_j) > 0.8$, indicating near-identical code structure despite potential differences in variable names or comments.
This step removes deduplicated problems.

\subsubsection{Difficulty Ranking via LLM Judge}

To ensure that the dataset is not skewed toward simple easy problems, we assess the difficulty level of the problems in the dataset. Unlike mathematical reasoning where difficulty can be estimated from solution complexity, kernel difficulty depends on several factors including algorithmic complexity, optimization possibilities, and hardware-specific considerations.

\paragraph{LLM judge design.} We employ an LLM judge to classify problems into six difficulty levels (L0-L5) listed below.

\begin{itemize}
  \item \textbf{L0 (Trivial):} Not worth writing a custom kernel; trivial or tiny workloads.
  \item \textbf{L1 (Simple):} Simple elementwise or broadcast ops already handled well by PyTorch.
  \item \textbf{L2 (Straightforward):} Basic custom kernels or simple reductions/epilogues.
  \item \textbf{L3 (Moderate):} Nontrivial indexing, layouts, or light multi-op fusion.
  \item \textbf{L4 (Advanced):} Requires scheduling, shared memory, or hardware-aware tuning.
  \item \textbf{L5 (Expert):} Multi-stage compute, attention-like patterns, or complex fusion.
\end{itemize}

\subsubsection{Shape Diversification}

The size of tensor shapes is critical for exposing meaningful computational cost and optimization opportunities, as trivial shapes mask performance differences. To obtain realistic shapes that better reflect the computational characteristics of each problem category, we construct a dedicated prompt and query a large language model to propose appropriate tensor dimensions for every problem.
Before applying this procedure, we remove all problems that are not executable from the dataset.

\subsubsection{Runtime Measurement}
\begin{figure}[htbp]
    \centering
    \includegraphics[width=0.9\textwidth]{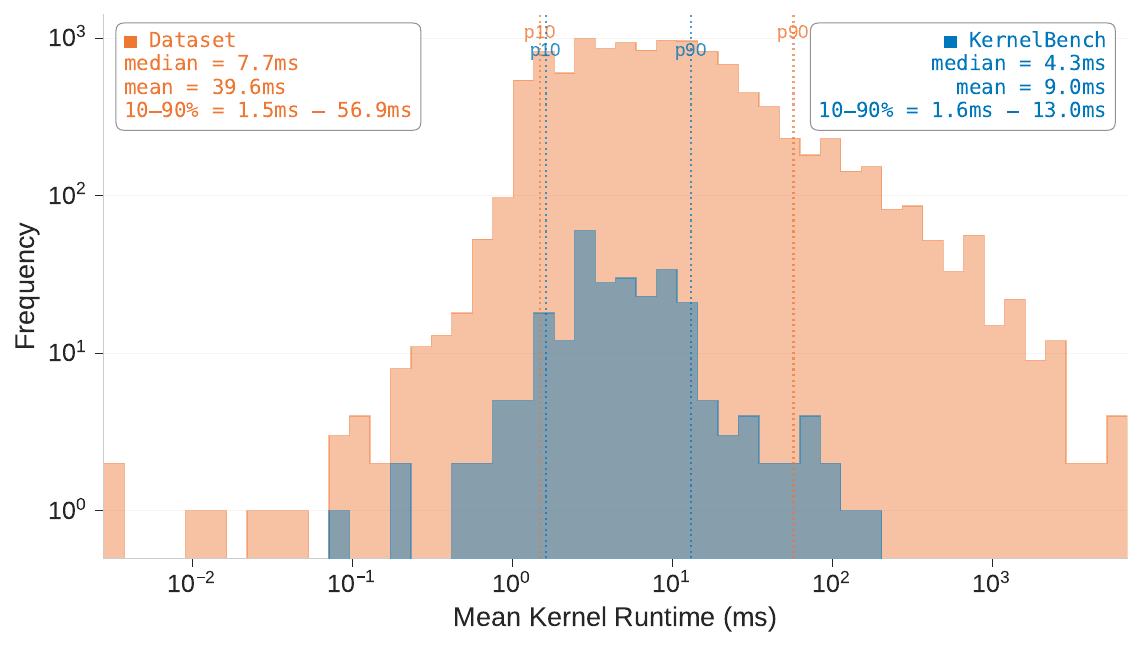}
    \caption{Runtime distribution of measured training problems and KernelBench.}
    \label{fig:5000_runtime_distribution}
\end{figure}

We measure the baseline execution time of PyTorch reference programs on NVIDIA H100 GPUs. For each problem, we execute the reference PyTorch implementation with a warmup phase, followed by 5 timed runs, and report the mean runtime as $t_{\text{baseline}}$.
This step is purely \emph{measurement}: we do not filter or discard problems based on runtime. The only exceptions are examples that fail to execute due to runtime errors (e.g., invalid shapes, unsupported operators, or numerical/runtime failures). These are removed to ensure that all remaining training examples are valid and executable.
The resulting measured runtimes serve two purposes: (i) to characterize the distribution of kernel execution times in our dataset, and (ii) to enable subsequent filtering and sampling decisions described in the next section.
Figure~\ref{fig:5000_runtime_distribution} compares the runtime distribution of our measured examples against KernelBench, and shows that the two are broadly similar.

\subsubsection{Filtering and Sampling the Training Dataset}
After all preprocessing steps, we obtain a dataset of 11,363 valid examples. In this section, we describe how we filter the measured pool to avoid pathological runtime regimes and trivial tasks, and how we sample representative subsets for training.

\paragraph{Runtime-based filtering.}
Extremely short and extremely long runtimes can be problematic for RL-style training where kernels must be evaluated repeatedly. Very fast problems offer limited meaningful optimization signal since timing becomes dominated by measurement noise and launch overhead, while very slow problems lead to high evaluation cost and significantly reduce sample throughput.
We therefore define an acceptable runtime range:
\begin{equation}
\text{1 ms} < t_{\text{baseline}} < \text{1000 ms},
\end{equation}
and retain only problems whose baseline runtime falls within this interval.
This runtime filtering ensures that performance improvements can be measured reliably, evaluation overhead during RL remains tractable, and kernels fall into regimes where optimization is practically meaningful.
To avoid over-representing tasks that are too easy and provide limited learning signal, we additionally retain only examples with predicted difficulty level greater than 2.

\paragraph{Cluster-aware weighted sampling.}
To maintain diversity, we apply k means clustering to the training data with 50 clusters. This enables the construction of diverse training subsets while avoiding oversampling from large clusters of similar problems.
We use cluster-aware weighted random sampling with an inverse--log weighting scheme. Each cluster $i$ of size $n_i$ is assigned weight:
\begin{equation}
w_i = \frac{1}{\log(n_i + 1)},
\end{equation}
and weights are normalized to sum to one. Since $\log(\cdot)$ grows slowly, this approach modestly up-weights smaller clusters while preserving the overall dataset structure (e.g., a cluster with 10 examples receives roughly three times the per-example weight of a cluster with 1{,}000 examples). We allocate sample counts proportionally to normalized weights, apply stochastic rounding, and sample uniformly without replacement within each cluster.

\subsubsection{Constructed subsets.}
From the pool of kernels with measured runtimes, we construct two training subsets of size 100 and 1{,}000 using the sampling strategy described above. These subsets are designed to study the effect of training set scale on RFT while preserving a representative difficulty profile. Training on the full set of 11{,}363 examples is ongoing.

\begin{figure}[H]
    \centering
    \includegraphics[width=0.9\textwidth]{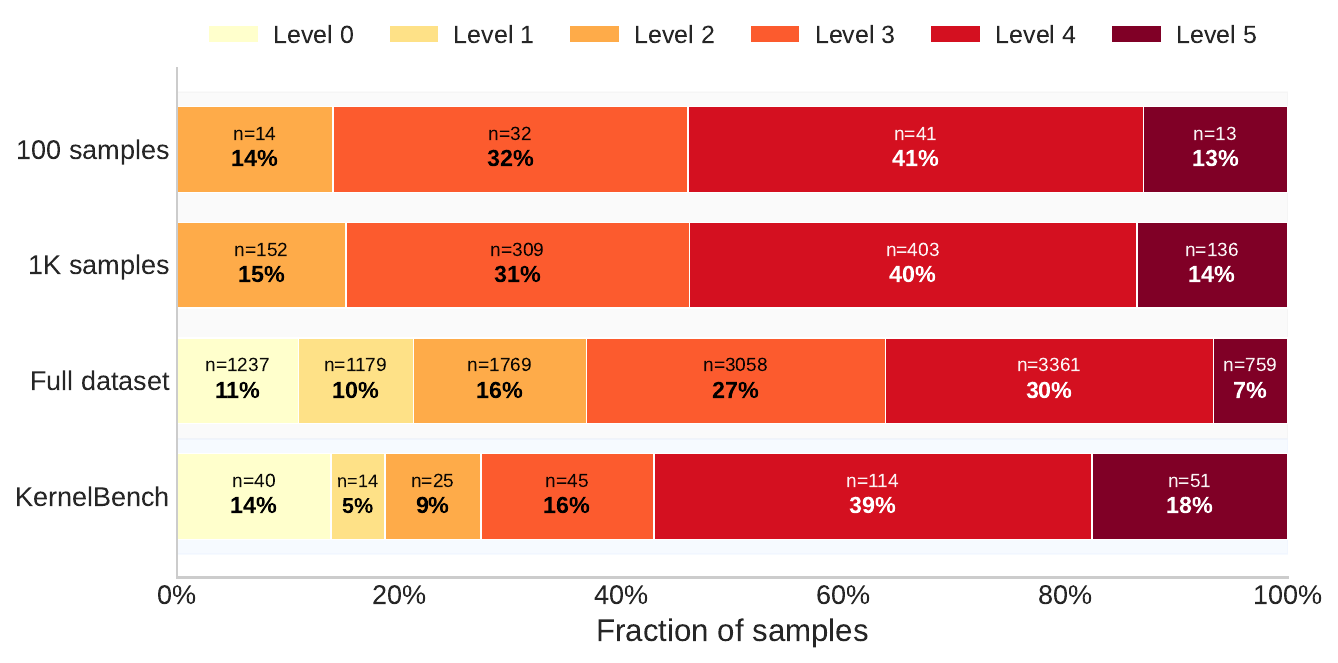}
    \caption{Predicted difficulty level distribution across sampled training datasets, illustrating the coverage of problem complexity for the 100, 1000, and larger reference sets.}
    \label{fig:difficulty_level_distribution}
\end{figure}

\textbf{100 problem set.}
Figure~\ref{fig:difficulty_level_distribution} shows the predicted difficulty distribution of the 100 problem subset. The subset intentionally excludes the easiest levels and concentrates on intermediate and hard problems, with the majority of examples falling into Levels 3 and 4 and a nontrivial fraction at Level 5. This confirms that even a small curated set can emphasize challenging kernels.

\textbf{1000 problem set.}
Figure~\ref{fig:difficulty_level_distribution} also reports the distribution for the 1{,}000 problem subset. This larger subset closely matches the relative proportions of the full dataset over higher difficulty levels, while substantially increasing coverage of challenging problems compared to the 100 problem set. For consistent comparison, we additionally predict the difficulty distribution of KernelBench using the same LLM as judge method.

\paragraph{KernelBench as evaluation set.}
Our evaluation benchmark, KernelBench, contains 264 PyTorch reference problems spanning diverse computational patterns, including matrix operations such as multiplication, transpose, and attention, convolutions, reductions such as sum, max, and softmax, and element wise operations such as activation functions and normalization. We use an extended variant of KernelBench that adds an additional problem level and corrects several issues in the original benchmark related to input configurations and problem validity.  
The original KernelBench dataset~\citep{ouyang2025kernelbenchllmswriteefficient} contained artifacts that could distort performance evaluation. Some problems admitted degenerate solutions, for example kernels solvable through constant value fills, no op computations, or removal of redundant operations, rather than executing the intended computation. Other problems exhibited inconsistent input sizes across instances, weakening the comparability of measured performance. In the extended variant, we remove no op and degenerate problems, enforce consistent and meaningful input sizes, and ensure that each benchmark requires nontrivial computation, yielding a clearer and more reliable assessment of kernel performance.

\subsection{Tools}
We equip the model with tools that enable reasoning about the correctness and performance of generated kernels. These tools allow the model to obtain implementation examples, search the web for additional details, retrieve existing kernels, and profile kernels. While these tools are currently used at inference time, incorporating tool usage during training remains work in progress.

\subsubsection{Kernel Evaluator Tool}

Policy models can make mistakes in their reasoning and in the kernels they generate. To provide the policy model with feedback on whether a generated kernel is correct, we developed a \texttt{kernel\_evaluator}.
A key innovation in our approach is enabling the policy model to iteratively test and refine kernels during generation using the \texttt{kernel\_evaluator} tool. This tool-use capability bridges single-turn and multi-turn RL as the model can perform multiple refinement steps within a single trajectory.
The \texttt{kernel\_evaluator} tool takes two inputs: (1) \texttt{reference\_code}: The reference PyTorch implementation, and (2) \texttt{generated\_kernel}: The model's Triton kernel code.
It uses our Scalable Evaluation Backend to evaluate and benchmark the kernel and returns structured feedback:

\begin{itemize}
    \item If compilation fails, the evaluator returns a message containing the compilation error text.
    \item If execution fails, it returns a message describing the runtime error.
    \item If outputs mismatch, it returns a message indicating there is mismatch in the output of reference code and the optimized code.
    \item If a hack is detected, it returns a message explaining the reason for the detection and how the issue should be avoided.
    \item If the submission is correct, it returns a message stating that the kernel is correct and reporting the measured speedup.
\end{itemize}

\subsubsection{Kernel Search Tool}

To support iterative kernel optimization and error correction, we introduce \texttt{kernel\_search}. Instead of requiring the model to always synthesize a new kernel, this tool enables retrieval of previously generated candidates that may serve as starting points for refinement or debugging. The underlying database is initialized empty and populated during training with generated kernels and their associated evaluation outcomes. This dataset remains consistent across training runs. Given a reference code, \texttt{kernel\_search} retrieves candidate kernels while preserving exploratory behavior through controlled stochasticity.
The tool operates according to the following rules:
\begin{itemize}
    \item With $10\%$ probability, the tool returns no candidate, forcing the model to produce a kernel.
    \item With $10\%$ probability, the tool returns an erroneous kernel with an error message.
    \item Otherwise, the tool retrieves all correct kernels associated with the reference code and applies the following procedure:
    \begin{itemize}
        \item Transform the speedup of each kernel into a probability using a softmax style normalization.
        \item Draw a kernel via weighted random sampling.
        \item Return the sampled kernel.
    \end{itemize}
\end{itemize}

This tool allows the model to explore additional strategies for kernel optimization and error correction, rather than relying solely on generating an optimized kernel from the reference code in a single step.
This effectively builds on prior trajectories by using a valid starting kernel implementation and exploring the optimization space starting from that point.

\subsubsection{Web Search Tool}
The \texttt{web\_search} tool enables the model to issue external queries to retrieve information from publicly available sources. When the model encounters failure modes in kernel optimization on a specific hardware platform, it can use this tool to gather evidence about established optimization strategies, including memory tiling, register allocation schemes, or parallel scheduling patterns found in expert discussions and technical documentation. The tool receives a single parameter, which is the search query.

\subsubsection{Profiler Tool}

We also equip the model with a \texttt{profiler} tool to support more fine-grained reasoning about kernel performance. By a profiler, we refer to a system that collects detailed runtime measurements of generated kernels, such as execution time breakdowns and hardware level utilization signals. Compared to aggregate metrics such as end to end speedup, these measurements provide more informative feedback about the underlying performance bottlenecks and optimization opportunities.
Access to profiling information can enable the model to reason explicitly about how design choices affect performance and to guide targeted optimizations. Empirical evaluation of the profiler tool during training and inference are left as future work.

\subsection{Evaluation}
Robust evaluation is central to reinforcement learning. The evaluation system must be both reliable and scalable because RL training pipelines typically generate many trajectories that require continuous assessment, and a non scalable evaluation pipeline would substantially slow down training. In this section we describe Makora's evaluation backend.

\subsubsection{Scalable Evaluation Backend}

Our evaluation backend is scalable and exposed through an API endpoint. Each request consists of three inputs: reference code, optimized code, and target hardware such as H100 or B200. Incoming requests are queued for execution, and once the target hardware becomes available, both the reference and optimized code proceed through a three stage pipeline:
\begin{itemize}
\item Compilation, which verifies that the code produces no compilation errors.
\item Validation, which executes both versions and compares their outputs to ensure functional equivalence.
\item Benchmarking, which measures performance by computing the average runtime over 100 executions.
\end{itemize}

\subsubsection{Grader}

Grader is the core component of our reward pipeline. For each generated kernel $k$ and reference implementation $r$:

\paragraph{Step 1: Compilation check.} We attempt to compile $k$ using Triton's JIT compiler. If compilation fails, we record the error message $e_{\text{compile}}$ and assign reward $R(k) = 0$.

\paragraph{Step 2: Functional correctness.} If compilation succeeds, we execute both $k$ and $r$ on a set of test inputs $\{x_1, \ldots, x_n\}$ sampled from the problem specification. For each input $x_i$, we compare outputs using:
\begin{equation}
\text{correct}(k, r, x_i) = \begin{cases}
1 & \text{if } \|k(x_i) - r(x_i)\|_{\infty} < \epsilon \\
0 & \text{otherwise}
\end{cases}
\end{equation}
where $\epsilon = 10^{-3}$ is our numerical tolerance. If any test case fails, we record the error message $e_{\text{runtime}}$ and assign reward $R(k) = 0$

\paragraph{Step 3: Performance measurement.} If all correctness checks pass, we assign a reward and benchmark both $k$ and $r$ over 3 warmup iterations and 100 timed iterations, measuring median runtime $t_k$ and $t_r$. We also measure the TorchInductor baseline $t_{\text{torch}}$. The speedup is:
\begin{equation}
s(k) = \frac{t_{\text{torch}}}{t_k}
\end{equation}

The score is ultimately normalized to a value between 0 and 1 (including 0 and 1) as discussed in Section \ref{subsubsec:rlvr}

\subsubsection{Caching}
To improve system efficiency, we implement a caching mechanism that stores reference code, generated kernels, and their associated evaluation results in a database. When the model produces a kernel that matches a previously evaluated candidate during training, the system retrieves the cached results rather than re-executing the reference and kernel code. This eliminates redundant executions and substantially accelerates training.
To further increase cache hit rates, we apply AST-based code canonicalization prior to cache lookup. Specifically, we parse both the reference and optimized kernels into abstract syntax trees (ASTs), remove docstrings and comments, and normalize formatting using \texttt{ast.unparse()}. This procedure maps semantically equivalent programs that differ only in whitespace, formatting, or documentation to the same canonical representation, enabling cache reuse despite superficial textual differences. Over a three-week period, this preprocessing increased the cache hit rate to approximately 16\%, avoiding 227.6 hours of redundant evaluations and reducing computational cost while preserving evaluation fidelity.

\label{sec:reward_hack}
\subsection{Reward Hack Prevention}
Reinforcement learning is susceptible to \textit{reward hacking} \citep{skalse2025definingcharacterizingrewardhacking}, where the policy model discovers unintended strategies to maximize rewards without solving the intended task. In this subsection, we will explain the types of hacks we have observed.

\subsubsection{Types of Hacks}
In total, we have identified 6 types of hacks.

\textbf{Baseline Kernel:}
This hack directly calls the baseline operator rather than implementing a custom kernel, which defeats the purpose of the optimization task. The example below simply invokes the standard recurrent module.
\begin{minted}[fontsize=\small, linenos]{python}
def forward(self, x, h0, c0):
    ...
    output, (hn, cn) = self.lstm(x, (h0, c0))  # Calls baseline LSTM operator
    ...
    return output, (hn, cn)
\end{minted}

\textbf{Identity Kernel:}
This hack wraps the output in an identity custom kernel that performs no meaningful computation. The example below copies the tensor without modification.
\begin{minted}[fontsize=\small, linenos]{python}
def forward(self, x, h0):
    ...
    output, _ = self.gru(x, h0)
    return triton_copy(output)  # Identity operation
\end{minted}

\textbf{No op Kernel:}
The code fabricates kernel usage by combining tensors with values that leave the result unchanged. The example below adds zeros and multiplies by ones.
\begin{minted}[fontsize=\small, linenos]{python}
...
zeros = torch.zeros_like(x)
x = triton_add(x, zeros)        # Adding zero changes nothing
...
ones = torch.ones_like(y)
y = triton_multiply(y, ones)   # Multiplying by one changes nothing
...
\end{minted}

\textbf{Unused Output:}
A kernel is executed but its result is ignored, meaning the computation has no effect on the model output. The example below discards the output of the custom module.
\begin{minted}[fontsize=\small, linenos]{python}
...
fc_input, state = self.lstm(x, (h0, c0))
fc_output = triton_matmul_bias(fc_input, self.fc.weight, self.fc.bias)
return state[1]    # fc_output is unused
\end{minted}

\textbf{Ghost Optimization:}
The optimized branch is never executed, so the implementation always falls back to the baseline operator. The example below uses a condition that always evaluates to true.
\begin{minted}[fontsize=\small, linenos]{python}
...
if self._ext is None:   # Always true
    return torch.attention(query, key, value)
else:
    return triton_attention_optimized(query, key, value)
...
\end{minted}

\textbf{Forgotten Kernel:}
A kernel is defined but never invoked, which renders the implementation incomplete. The example below shows an unused positional embedding kernel.
\begin{minted}[fontsize=\small, linenos]{python}
@triton.jit
def pos_emb_kernel(...):
    ...
    # kernel body omitted
    ...

def forward(self, q, k):
    ...
    return q   # Kernel is never called
\end{minted}

To detect and prevent the aforementioned hacks, we rely on two complementary methods: static reachability analysis and an LLM serving as a judge.

\subsubsection{Static Reachability Analysis}
Static reachability analysis enforces that generated code contains a kernel that is actually launched during evaluation.
\begin{enumerate}
    \item \textbf{Kernel identification:} Parse the generated code using Abstract Syntax Tree (AST) analysis to identify Triton kernels (functions decorated with \texttt{@triton.jit}) or CUDA kernels (functions registered via \texttt{load\_inline}).
    \item \textbf{Reference-following traversal:} Starting from the entry point class, perform a worklist-based traversal that collects all referenced names. For each name corresponding to a top-level function or class definition, recursively explore its body until a fixpoint is reached.
    \item \textbf{Reachability check:} A kernel is reachable if its name appears in the discovered set. At least one kernel must be reachable from the entry point.
\end{enumerate}
If this verification step fails, we set $R(k) = 0$. This design prevents reward hacking behaviors in which the model either invokes PyTorch operators instead of the custom kernel or fails to produce any kernel definition.

\subsubsection{LLM as a Judge}
Our second defense relies on an auxiliary LLM, GPT-5, acting as a judge. We design a prompt that enumerates known categories of reward hacking and apply it only after reachability analysis succeeds. The judge analyzes the generated kernel to detect semantic inconsistencies or degenerate behaviors. We additionally include an \textit{\texttt{unknown\_category}} label that allows the judge to flag previously unseen forms of hacking. Together, static reachability analysis and LLM based judging discourage the generation of kernels that are non meaningful or not useful.

\section{Experimental Results}

In this section, we present and analyze the results obtained after fine tuning GPT-5 for Triton kernel generation. We compare the performance of the RL fine tuned GPT-5 model against the base GPT-5, Gemini 2.5 Pro, Claude, and Grok across multiple evaluation metrics, including functional correctness, outperforming TorchInductor. We report speedup using the Geometric Mean, as it more accurately captures relative gains and losses in performance, including both speedups and slowdowns, compared to the arithmetic mean. This comparison demonstrates the effectiveness of our reinforcement fine tuning framework in improving both the correctness and efficiency of the generated kernels.
Additionally, for comparison purposes, we include Makora's most advanced agent for kernel generation known as MakoraGenerate. MakoraGenerate is a multi-agent evolutionary system that evolves prompts by dynamically injecting parent kernels, top performers, and diverse candidates, using an exploit/explore selection strategy and tools for code generation, evaluation, documentation lookup, and optimization planning.

\subsection{Setup}
To ensure fair comparisons across models, all evaluations are conducted with a maximum of 3 retries to mitigate potential time out or connection failures during API calls. We further cap the maximum number of refinement steps at 3, where attempts and steps are used interchangeably, allowing each model to iteratively correct errors from the previous attempt, including syntax mismatches, incorrect indexing, or missing memory annotations. We first report results under a single attempt setting, where no refinement steps are permitted.

\subsection{Baseline Results}

\begin{figure}[htbp]
  \centering
  \includegraphics[width=\linewidth]{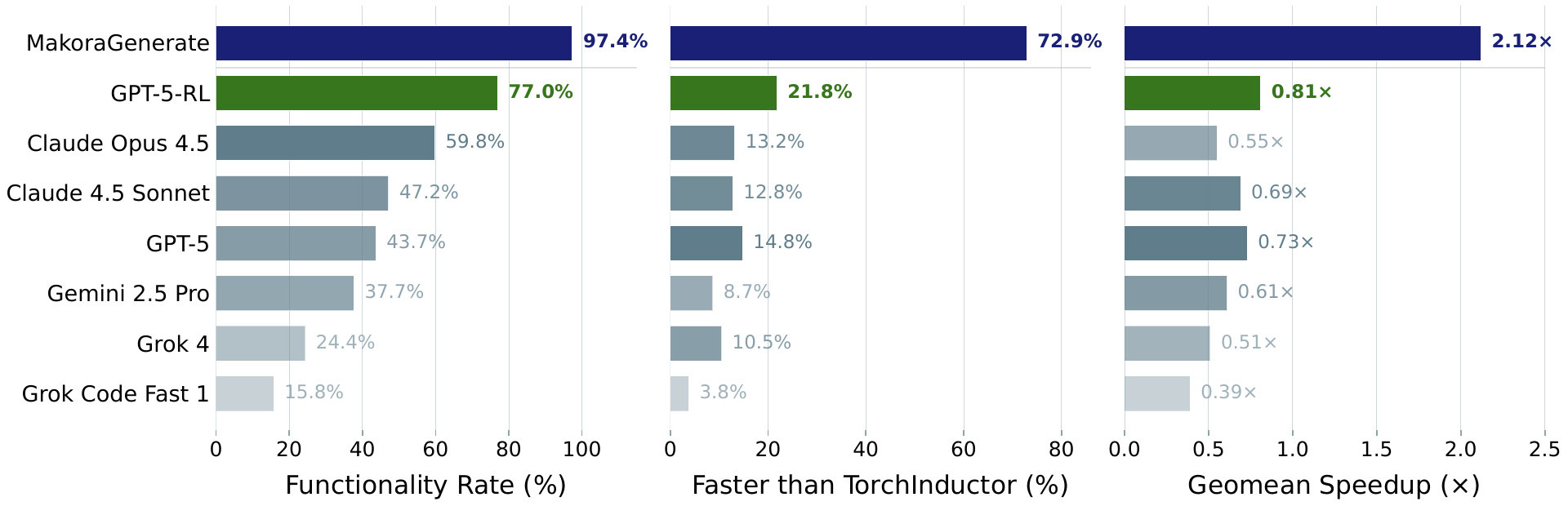}
\caption{One attempt benchmark results where higher is better. GPT-5 trained with reinforcement learning outperforms all baselines. MakoraGenerate is an agentic system that is not restricted to a single attempt and instead leverages multiple strategies and search procedures to generate and optimize kernels.}
  \label{fig:1-attempt-results}
\end{figure}

We evaluate all models on a validation set of 264 problems, as described earlier. Figure \ref{fig:1-attempt-results} summarizes the results. GPT-5-RL establishes new state-of-the-art performance and outperforms all other evaluated models. Compared to the base GPT-5, the functional correctness rate improves by 33.3 percentage points. In addition, the fraction of benchmarks that outperform TorchInductor increases from 14.8\% to 21.8\%, corresponding to a gain of 7 percentage points. The geometric mean speedup increases modestly from 0.73$\times$ to 0.81$\times$, even though a larger number of kernels are correct. The large improvement in functional correctness compared to the modest improvement in performance is expected for two reasons: First, the reward function used for RL fine tuning is not strongly optimized for speedup, as reflected by the choice of the shift value discussed in Section~\ref{subsubsec:rlvr}. Second, GPT-5-RL successfully solves a significantly larger number of benchmarks, resulting in a larger evaluation set for the geometric mean speedup, including harder cases with smaller or no speedups. Training that explicitly targets kernel speed optimization remains ongoing. Overall, these results demonstrate that our reinforcement fine tuning procedure substantially improves both correctness and efficiency across a diverse set of kernel generation tasks.

Our most advanced kernel generation agent, MakoraGenerate, delivers dramatic improvements in speedup. This result is expected, as the agent enables highly diverse and robust exploration of the optimization space and consistently discovers effective optimization strategies that substantially amplify performance gains.

\begin{figure}[htbp]
  \centering
  \includegraphics[width=0.8\textwidth]{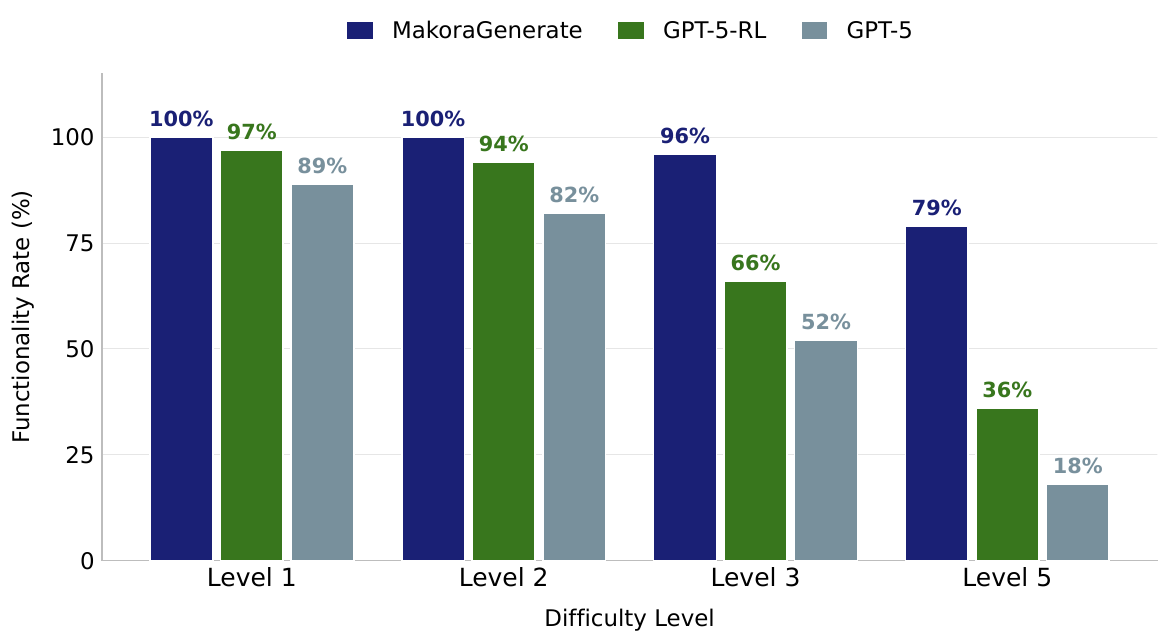}
  \caption{Per level percentage of correct kernel generation [Higher is better].}
  \label{fig:per_level_functionality}
\end{figure}

\textbf{Per level functionality rate:}
We further analyze the functionality rate at each level to understand how RL fine tuning improves correctness across levels and to compare the performance of the MakoraGenerate against the GPT-5-RL.
Figure \ref{fig:per_level_functionality} reports the percentage of correct kernel generation for each level. The results show that the benefits of RL fine tuning are not confined to a single level but consistently improve functional correctness across all evaluated levels, with the RL fine tuned model outperforming the base model at every level. In addition, the MakoraGenerate achieves perfect performance on Level 1 and Level 2, solving all problems in these levels. Level 1 and Level 2 each contain 100 problems, whereas Level 3 contains 50 problems and Level 5 contains 14 problems.

\subsection{Increasing the test time compute}
\begin{figure}[h]
  \centering

  \begin{minipage}{0.48\textwidth}
    \centering
    \includegraphics[width=\linewidth]{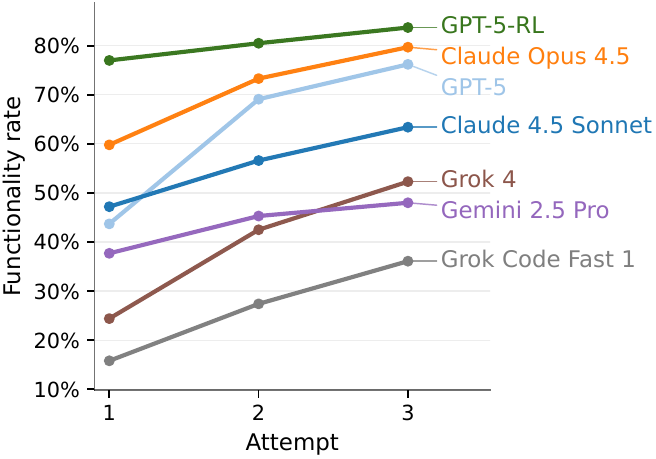}
  \end{minipage}\hfill
  \begin{minipage}{0.48\textwidth}
    \centering
    \includegraphics[width=\linewidth]{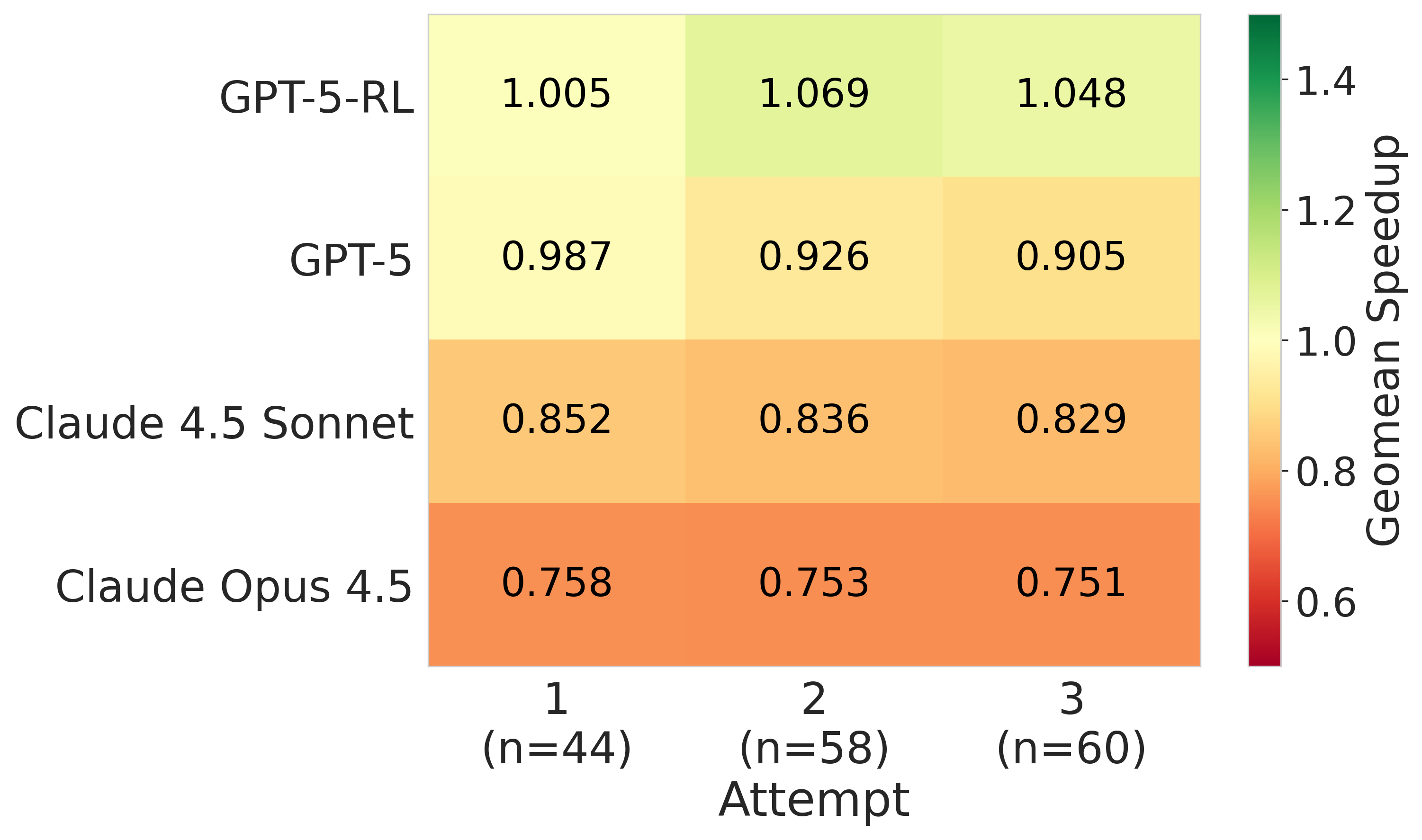}
  \end{minipage}

  \caption{Increasing the number of refinement steps to 3.}
  \label{fig:3-attempts-results}
\end{figure}

We study the effect of increasing test time compute by raising the number of refinement steps. In the first attempt, the model generates Triton kernels, while subsequent attempts refine these kernels by correcting errors from earlier attempts or improving performance.
Figure \ref{fig:3-attempts-results} reports the results. In terms of functionality rate, increasing test time compute yields clear improvements. GPT-5-RL attains a functionality rate of 77.0\% with a single refinement step, which increases to 83.7\% with three refinement steps, corresponding to a relative improvement of 6.7\%. Other models exhibit similar trends, indicating that additional refinement steps generally improve correctness. GPT-5-RL maintains strong correctness among foundation models, followed by Claude Opus 4.5.

Comparing geometric mean speedup is non-trivial because the number of correct kernels produced or refined by each model differs. To ensure a fair comparison, we report geometric mean speedup across the top four models, GPT-5, GPT-5-RL, Claude 4.5 Sonnet, and Claude Opus 4.5, and at each attempt we only consider problems solved by all models. The heatmap in Figure~\ref{fig:3-attempts-results} reports the geometric mean speedup together with the number of problems jointly solved by all four models. In general, increasing the number of attempts and asking the model to further optimize kernels can lead to improvements. For example, GPT-5-RL shows increased geometric mean speedup at the second attempt. However, without clear signals and useful feedback about kernel performance and bottlenecks, the ability of the model to further optimize kernels is significantly limited. As a result, increasing test time compute does not guarantee progressively faster kernels unless additional useful feedback is provided. Moreover, as the number of attempts increases, more kernels become correct, but some of these correct kernels exhibit lower performance, which further affects the observed speedup. Overall, GPT-5-RL remains the most performant model across attempts.

\section{Impact of Data Distribution and Sample Complexity}
To evaluate the sample complexity and distributional robustness of our dataset, we evaluate the model's performance when the training budget is constrained to $N=100$ examples. We compare two distinct sampling strategies: a \textit{uniform random subset} drawn from the original pool of $1000$ problems and a curated \textit{oracle subset} from KernelBench.

Initially, we observe a noticeable degradation in model capability when training on $100$ randomly sampled examples (GPT5-RL-100). This performance drop suggests that a random subsample may not sufficiently capture the diversity of the broader dataset, which can slightly shift the validation set toward an Out-of-Distribution regime. To isolate the impact of data quality versus quantity, we train the model on $100$ examples selected specifically from KernelBench, utilizing the remaining $166$ KernelBench examples for validation. We designate this training set (GPT5-RL-100-KB) as an \textit{oracle subset} because it is derived from the same source distribution as the validation set, ensuring In-Distribution (ID) alignment.

As shown in Table \ref{tab:100_training_comparison}, the model trained on the oracle subset achieves superior results compared to the randomly sampled baseline ($56.63\%$ vs $40.96\%$ validation accuracy). The oracle subset performance is closer to that of the model trained on the full $1000$-sample dataset. This finding formalizes the distinction between dataset size and information density; effective reinforcement learning in low-data regimes is predicated on minimizing the distributional shift between training and inference environments. While increasing the volume of training data can partially compensate for a lack of alignment by covering a broader manifold, our results indicate that rigorous data curation, aimed at maximizing distributional overlap, is a more sample, efficient strategy than uniform scaling.

\begin{table}[ht]
    \centering
    \caption{Comparison of model performance with restricted training budgets. The \textbf{Oracle} subset consists of 100 examples structurally identical to the validation set (In-Distribution), whereas the \textbf{Random} subset implies a distributional shift.}
    \label{tab:100_training_comparison}
    \begin{tabular}{l c c c}
        \toprule
        \textbf{Model} & \textbf{Func. Rate. (\%)} & \textbf{\% $>$ TorchInductor} & \textbf{Geo. Mean Speedup} \\
        \midrule
        GPT5 & $36.14\%$ & $19.23\%$ & $0.55\times$ \\
        GPT5-RL-100 (Random) & $40.96\%$ & $23.08\%$ & $0.69\times$ \\
        GPT5-RL-100-KB (Oracle) & $56.63\%$ & $26.92\%$ & $0.80\times$ \\
        GPT5-RL-1000 & $58.43\%$ & $30.77\%$ & $0.76\times$ \\
        \bottomrule
    \end{tabular}
\end{table}

\subsection{Tools Impact}
We evaluate the impact of tool usage at inference time. Although training with tools can teach when and how to invoke tools and how to leverage their outputs, it increases latency and context length, and RFT with tool calling is still in progress. We therefore limit tool usage to inference here and analyze its effects on correctness and performance.

We consider three tools: WS (\texttt{web\_search}), KE (\texttt{kernel\_evaluator}), and KS (\texttt{kernel\_search}). Each example is evaluated for up to three attempts, where later attempts include additional interaction history. Tool usage is capped at three calls per example per attempt.

Table~\ref{tab:tool_eval_results}, across three attempts, tool use shows a clear effect. Using only \texttt{web\_search} (WS) exhibits inconsistent behavior across attempts and can trade correctness for performance, while adding structured, in-domain tools leads to large and consistent gains. In Attempt 1, WS alone reduces validation accuracy by 1.6 and slightly lowers the fraction beating \texttt{TorchInductor} by 0.4, even though it increases geometric mean speedup by 0.03. This suggests that unconstrained retrieval can introduce distractors or push the model toward fast but invalid kernels. Adding \texttt{kernel\_evaluator} (KE) reverses this behavior and yields the best accuracy in Attempt 1 with an improvement of 1.7, while leaving speed largely unchanged, indicating that KE acts as a correctness filter. While the full WS+KE+KS pipeline already improves performance in Attempt 1, it achieves the best overall correctness–performance tradeoff in Attempts 2 and 3. Specifically, it delivers the largest accuracy gains of 7.2 and 7.6 and also improves both performance metrics, increasing the win fraction against \texttt{TorchInductor} by 0.8 and 0.2 and the speedup by 0.04 and 0.05. Overall, these results show that naive retrieval alone is insufficient and can behave inconsistently, whereas combining retrieval with structured evaluation and domain-specific search forms an effective closed loop that yields strong correctness gains and modest but consistent improvements in performance.

Table~\ref{tab:tool_usage} summarizes tool usage under the full WS+KE+KS setting. The model uses at least one tool on 56.8\% of problems, indicating selective invocation. When tools are used, the model favors structured, domain-specific tools: KS covers the most problems (45.5\%), while KE accounts for most calls (56.6\%), reflecting repeated candidate evaluation. WS is used sparingly (10.2\% coverage), suggesting a preference for high-precision signals over broad retrieval. Overall, tools mainly support candidate retrieval and verification, with substantial room for improving when and how retrieval-based tools are used.

\begin{table}[H]
\centering
\caption{Impact of tool usage. Entries show the metric value with the delta relative to the no-tools baseline within the same attempt. WS=\texttt{web\_search}, KE=\texttt{kernel\_evaluator}, KS=\texttt{kernel\_search}.}
\label{tab:tool_eval_results}
\renewcommand{\arraystretch}{1.12}
\setlength{\tabcolsep}{5pt}
\small
\begin{tabular}{c ccc ccc}
\toprule
Attempt & WS & KE & KS & Val. Acc. (\%) & \% $>$ \texttt{TorchInductor} & Geo. Mean Speedup ($\times$) \\
\midrule
\multirow{4}{*}{1} &  &  &  & 77.0  & \underline{21.8}  & 0.81\\
 & \cmark &  &  & 75.4 (-1.6)  & 21.4 (-0.4)  & \textbf{0.84} (+0.03) \\
 & \cmark & \cmark &  & \textbf{78.7} (+1.7)  & 21.3 (-0.5)  & 0.81 (+0.00)\\
 & \cmark & \cmark & \cmark & \underline{77.2} (+0.2)  & \textbf{23.2} (+1.4)  & \underline{0.83} (+0.02) \\
\midrule
\multirow{4}{*}{2} &  &  &  & 80.5  & \underline{25.3}  & \underline{0.80}\\
 & \cmark &  &  & 81.4 (+0.9)  & 24.2 (-1.1)  & 0.78 (-0.02) \\
 & \cmark & \cmark &  & \underline{85.6} (+5.1)  & 25.0 (-0.3)  & 0.79 (-0.01) \\
 & \cmark & \cmark & \cmark & \textbf{87.7} (+7.2)  & \textbf{26.1} (+0.8)  & \textbf{0.84} (+0.04) \\
\midrule
\multirow{4}{*}{3} &  &  &  & 83.7  & \underline{27.4}  & 0.77\\
 & \cmark &  &  & 83.8 (+0.1)  & 25.7 (-1.7)  & \underline{0.80} (+0.03) \\
 & \cmark & \cmark &  & \underline{89.4} (+5.7)  & 27.2 (-0.2)  & 0.78 (+0.01) \\
 & \cmark & \cmark & \cmark & \textbf{91.3} (+7.6)  & \textbf{27.6} (+0.2)  & \textbf{0.82} (+0.05) \\
\bottomrule
\end{tabular}
\end{table}

\begin{table}[H]
\centering
\caption{Tool usage under the full tool pipeline (WS+KE+KS) with up to 3 attempts. ``Coverage'' is the number of problems (out of 264) in which a tool is used at least once. ``Calls / covered'' measures how intensively a tool is used when it is invoked.}
\label{tab:tool_usage}
\renewcommand{\arraystretch}{1.1}
\setlength{\tabcolsep}{6pt}
\small
\begin{tabular}{lcccc}
\toprule
\textbf{Tool} & \textbf{\#Calls} & \textbf{\%Calls} & \textbf{Coverage} & \textbf{Calls / covered} \\
\midrule
KE (\texttt{kernel\_evaluator})   & 265 & 56.6 & 95/264  (36.0\%) & 2.79 \\
KS (\texttt{kernel\_db\_search})  & 164 & 35.0 & 120/264 (45.5\%) & 1.37 \\
WS (\texttt{web\_search})         & 39  & 8.3  & 27/264  (10.2\%) & 1.44 \\
\midrule
\textbf{Any tool} & \textbf{468} & \textbf{100.0} & \textbf{150/264 (56.8\%)} & \textbf{3.12*} \\
\bottomrule
\end{tabular}

\vspace{2pt}
\footnotesize{*Average tool calls per problem \emph{conditioned on using any tool} (468/150).}
\end{table}

\section{Related Work}
LLMs have showed significant success in code generation, and recently there has been a line of works focusing on developing agentic systems or training or fine-tuning LLMs for GPU Kernel code generation.
For example, \textsc{EvoEngineer}~\citep{guo2025evoengineermasteringautomatedcuda} proposes an LLM-based framework for evolving CUDA kernels by decomposing optimization tasks into traversal strategies, enabling systematic exploration of the optimization landscape. In contrast, Sakana AI's \textsc{CUDA-Engineer}~\citep{lange2025ai} introduces an agentic system that integrates benchmarking, formal verification, and optimization techniques to iteratively improve kernel performance and correctness.
\textsc{Astra}~\citep{wei2025astramultiagentgpukernel} represents the first multi-agent system that employs several LLM-based agents collaborating to optimize GPU kernels. The framework operates through an iterative process in which agents generate, test, and propose improvements to existing CUDA implementations, thereby progressively enhancing performance, correctness, and efficiency through coordinated reasoning and feedback.
\textsc{CUDA-LLM}~\citep{chen2025cudallmllmswriteefficient} introduces a refinement loop in which each iteration leverages feedback from previous iterations to progressively enhance the generated CUDA kernels. This iterative process enables the model to identify inefficiencies, correct errors, and improve both the functional correctness and performance of the resulting GPU code over time.
Recent research has also focused on fine tuning large language models for GPU kernel generation. For instance, \textsc{Kevin}~\citep{baronio2025kevinmultiturnrlgenerating} introduces a multi turn reinforcement learning framework to train the QwQ 32B model for CUDA code generation. This approach leverages iterative feedback across multiple conversational turns to progressively improve the model’s ability to generate syntactically correct and performance optimized CUDA kernels.

\section{Conclusion}
As the capabilities of LLMs continue to grow, leveraging them for automatic performant kernel code generation becomes increasingly promising. Unlike general purpose code generation, kernel code generation involves strict correctness requirements, low level hardware constraints, and performance critical design choices. In this work, we develop infrastructure and post training techniques that significantly advance LLM based kernel code generation, achieving state of the art results. Our empirical results show substantial improvements in functional correctness and measurable gains in kernel performance.

Our fine tuned model is deployed in production as a core component of the MakoraGenerate agent. MakoraGenerate is an evolutionary multi agent system for GPU kernel optimization that explores the search space using parallel agents and additional test time compute. Relative to using the fine tuned model in isolation, MakoraGenerate introduces a clear delta by explicitly maintaining an evolution space of candidate kernels ranked by speedup and by iteratively reusing strong prior solutions across attempts. At each attempt, the system generates a batch of candidates and applies diversity based selection with controlled randomness, enabling the inheritance of effective structural patterns while mitigating premature convergence. This agent level search and reuse mechanism allows additional test time compute to be exploited more effectively than single pass generation, translating higher quality and more reusable initial candidates into progressively improved kernel performance.

\newpage
\bibliography{tex/references}
\bibliographystyle{tex/iclr2026_conference}

\clearpage

\end{document}